\documentclass[journal]{IEEEtran}

\usepackage{booktabs}
\usepackage{threeparttable}
\usepackage{tablefootnote}
\usepackage{pifont}
\usepackage[utf8]{inputenc}
\usepackage{amsmath,amsfonts}
\usepackage{cite}
\usepackage{multicol}
\usepackage{graphicx}
\usepackage{array}
\usepackage{multirow}
\usepackage{graphicx}
\usepackage{subcaption}
\usepackage{enumitem}
\usepackage{color}
\usepackage{nomencl}
\usepackage{etoolbox}
\usepackage{amssymb}
\usepackage{comment}
\usepackage{mathtools}
\usepackage{yfonts}
\usepackage{soul}
\usepackage{hyperref}
\usepackage{dsfont}
\usepackage[acronym]{glossaries}
\usepackage{nomencl}
\usepackage{subfiles}
\usepackage[utf8]{inputenc}
\usepackage{mathtools}
\usepackage{amsmath}
\usepackage{tikz}
\usepackage{algorithm}
\usepackage{algpseudocode}
\usepackage{etoolbox}
\usepackage{fnpct}
\usepackage[font=small]{caption}
\usepackage[font=small]{subcaption}           
\ifCLASSOPTIONcompsoc \usepackage[caption=false,font=normalsize,labelfon
t=sf,textfont=sf]{subfig}
\else
\usepackage{tikz}

\ifCLASSINFOpdf

\else

\fi

\begin{document}

\title{Ramping-aware Enhanced Flexibility Aggregation of Distributed Generation with Energy Storage in Power Distribution Networks}

\author{
\IEEEauthorblockN{Hyeongon Park\thanks{School of Systems Management and Safety Engineering, Pukyong National University, Busan, 48516, South Korea (e-mail:\ hyeongon@pknu.ac.kr)}, 
Daniel K. Molzahn\thanks{School of Electrical and Computer Engineering, Georgia Institute of Technology, Atlanta, GA, 30313, USA (e-mail:\ molzahn@gatech.edu)}, and 
Rahul K. Gupta\thanks{School of Electrical Engineering and Computer Science, Washington State University, Pullman, WA, 99163, USA (e-mail:\ rahul.k.gupta@wsu.edu)}}
}
\makeatletter
\patchcmd{\@maketitle}
  {\addvspace{0.5\baselineskip}\egroup}
  {\addvspace{-2\baselineskip}\egroup}
  {}
  {}
\makeatother
\maketitle

\begin{abstract}
Power distribution networks are increasingly hosting controllable and flexible distributed energy resources (DERs) that, when aggregated, can provide ancillary support to transmission systems. However, existing aggregation schemes often ignore the ramping constraints of these DERs, which can render them impractical in real deployments.
This work proposes a ramping-aware flexibility aggregation scheme, computed at the transmission–distribution boundary, that explicitly accounts for DER ramp limits and yields flexibility envelopes that are provably disaggregable. To further enhance the attainable flexibility region, we introduce a novel “pre-ramping” strategy, which proactively adjusts resource operating points to enlarge the aggregated flexibility envelope while preserving both network feasibility and disaggregation guarantees. 
The proposed method demonstrates a 5.2\% to 19.2\% improvement in flexibility relative to the baseline model, depending on system conditions. We validate the scheme on an IEEE-33 bus distribution system and provide formal proofs showing that both aggregation strategies are disaggregable for all feasible trajectories within the aggregate flexibility envelope.
\end{abstract}


\IEEEpeerreviewmaketitle

\section*{Nomenclature}

\textit{Notation:} Superscripts $\wedge$ and $\vee$ denote variables associated with the upper and lower envelope trajectories, respectively. The superscript ``$\bullet$'' indicates quantities evaluated at the pre-ramped operating point.

\vspace{0.5em}

\subsubsection*{Indices}
\begin{IEEEdescription}[\IEEEusemathlabelsep\IEEEsetlabelwidth{$t \in \mathcal{T}$}]
\item[$t \in \mathcal{T}$] Time index.
\item[$g \in \mathcal{G}$] Generator index.
\item[$e \in \mathcal{E}$] Energy storage system (ESS) index.
\item[$n \in \mathcal{N}$] Node index.
\item[$\ell \in \mathcal{L}$] Line index.
\item[$s$] Envelope index, $s \in \{\wedge,\vee\}$.
\end{IEEEdescription}

\vspace{0.5em}

\subsubsection*{Parameters}
\begin{IEEEdescription}[\IEEEusemathlabelsep\IEEEsetlabelwidth{$P^{\mathrm{GCP}}_{\wedge,t},PP$}]
\item[$\Delta t$] Duration of a time step.
\item[$P^{\min}_g, P^{\max}_g$] Active power limits of generator $g$.
\item[$Q^{\min}_g, Q^{\max}_g$] Reactive power limits of generator $g$.
\item[$R^{\uparrow}_g$, $R^{\downarrow}_g$] Ramp-up/down limits of generator $g$ per time step.
\item[$P^{\max}_e$] Maximum charging/discharging power output of ESS $e$.
\item[$E^{\min}_e$, $E^{\max}_e$] State-of-charge (SoC) limits of ESS $e$.
\item[$E^{\mathrm{init}}_e$] Initial SoC of ESS $e$.
\item[$P^{\mathrm{init}}_g$] Initial active power output of generator $g$.
\item[$\kappa_e$] Energy efficiency factor of ESS $e$.
\item[$\boldsymbol{p}^{\text{load}}_t,\boldsymbol{q}^{\text{load}}_t$] Nodal active and reactive power demand.
\item[$\boldsymbol{p}^{\text{PV}}_t$] Nodal PV generation.
\item[$\underline{\boldsymbol{u}}$, $\overline{\boldsymbol{u}}$] Lower and upper bounds on squared nodal voltages.
\end{IEEEdescription}

\vspace{0.5em}

\subsubsection*{Decision Variables}
\begin{IEEEdescription}[\IEEEusemathlabelsep\IEEEsetlabelwidth{$\boldsymbol{p}^{\mathrm{inj},s\bullet}_tPP$}]

\item[$p^{\mathrm{GCP},s}_{t}$] Net active power exchanged at the grid connection point (GCP) for envelope $s$.
\item[$p^{\mathrm{GCP},s\bullet}_{t}$] Pre-ramped net active power at the GCP for envelope $s$.
\item[$p^{s}_{g,t}, q^{s}_{g,t}$] Active and reactive power outputs of generator $g$ for envelope $s$.
\item[$p^{s}_{e,t}$] Active power output of ESS $e$ for envelope $s$ (positive for discharging).
\item[$e^{s}_{e,t}$] SoC of ESS $e$ for envelope $s$.
\item[$p^{\mathrm{pre},s}_{g,t}$] Generator pre-ramp variable for envelope $s$.
\item[$p^{\mathrm{pre},s}_{e,t}$] ESS pre-ramp variable for envelope $s$.
\item[$e^{s\bullet}_{e,t}$] Pre-ramped SoC of ESS $e$ for envelope $s$.
\item[$\boldsymbol{u}^{s}_t$] Nodal squared voltages for envelope $s$.
\item[$\boldsymbol{u}^{s\bullet}_t$] Pre-ramped nodal squared voltages for envelope $s$.
\item[$\boldsymbol{p}^{\mathrm{inj},s}_t, \boldsymbol{q}^{\mathrm{inj},s}_t$] Nodal active and reactive power injections for envelope $s$.
\item[$\boldsymbol{p}^{\mathrm{inj},s\bullet}_t$] Pre-ramped nodal active power injections for envelope $s$.
\end{IEEEdescription}

\section{Introduction}

\nocite{yurdakul2024flexible,luo2025uncertainty,li2025aggregate,gupta2025grid,
wen2022aggregate,chen2019aggregate,ozturk2024alleviating,wen2025quantifying,
chen2021leveraging,al2024efficient,na2024aggregating,
gooding2024microsoft,cha2025potential,alipour2019real,michaelson2021review,
huang2023research}

\begin{table*}[t]
\centering
\begin{threeparttable}
\caption{Comparison of Existing Studies on Flexibility Aggregation}
\label{tab:related_work}
\renewcommand{\arraystretch}{1.2}
\newcolumntype{P}[1]{>{\centering\arraybackslash}p{#1}}
\newcolumntype{L}[1]{>{\raggedright\arraybackslash}p{#1}}
\begin{tabular}{L{2.2cm} p{3.1cm} L{2.1cm} P{1.0cm} P{1.0cm} L{1.8cm} p{3.9cm}}
\hline
\textbf{Reference} & \textbf{Objective} & \textbf{Grid Model} & \textbf{Explicit Ramp}\tnote{$\dagger$} & \textbf{Pre-ramp} & \textbf{Disaggregation} & \textbf{Remarks} \\
\hline

Wen et al.~\cite{wen2022aggregate}
& Exact derivation of temporally coupled aggregate flexibility
& Linearized AC / shift-factor model
& \ding{55}& \ding{55}
& Guaranteed
& Exact but relies on a complex projection framework that is hard to extend for additional constraints.\\

Chen et al.~\cite{chen2019aggregate}
& Maximization of inner-box flexibility
& Linearized multiphase
& \ding{55} & \ding{55}
& Guaranteed
& Inner-box construction ensures disaggregation. \\

Öztürk et al.~\cite{ozturk2024alleviating}
& Constructing an efficient inner approximation of the Minkowski sum 
& None
& \ding{55}
& \ding{55}
& Guaranteed
& Constructs a scalable inner approximation of storage flexibility using a vertex-based method.\\

Wen et al.~\cite{wen2025quantifying}
& Flexibility cost/value quantification for TSO–DSO coordination
& LinDistFlow
& \ding{55} & \ding{55}
& Guaranteed
& Uses inner-approximated flexibility to derive time-coupled cost models and marginal prices. \\

Huang et al.~\cite{huang2023research}
& Minimization of infeasibility via boundary shrinkage
& Linearized AC/DC model
& \ding{51} & \ding{55}
& Not addressed
& Initializes a  flexibility region and iteratively shrinks it by removing infeasible trajectories. \\

\textbf{Ramping-aware baseline model} (Section~\ref{sec:Old_Formulation})
& Maximization of inner-box flexibility
& LinDistFlow
& \ding{51} & \ding{55}
& Guaranteed
& Time-coupled envelope under DER and grid constraints; no pre-ramping. \\

\textbf{Pre-ramped enhanced model }(Section~\ref{sec:New_Formulation})
& Maximization of inner-box flexibility
& LinDistFlow
& \ding{51} & \ding{51}
& Guaranteed
& Pre-ramping enlarges the feasible envelope while preserving DER and grid constraints. \\

\hline
\end{tabular}

\begin{tablenotes}
\item[$\dagger$] “Explicit ramp” denotes direct limits on power change rates 
(as in generator ramping). DER models with only state-dependent time coupling 
(e.g., ESS SoC) are state-coupled rather than explicitly ramp-limited.
\end{tablenotes}
\end{threeparttable}
\vspace{-1.5em}
\end{table*}

The increasing penetration of renewable energy resources has significantly heightened the need for operational flexibility in modern power systems to reliably balance variability and uncertainty. This stochastic nature often leads to frequency instability and acute shortages in short-term balancing, frequently triggering price spikes in real-time markets \cite{yurdakul2024flexible}. In response, system operators such as California Independent System Operator (CAISO), Midcontinent Independent System Operator (MISO), and Southwest Power Pool (SPP) have established and operated markets for Flexible Ramping Products (FRPs) \cite{luo2025uncertainty}. Within the FRP market framework, generation resources are compensated for providing upward and downward ramping capability, enabling them to strategically offer flexibility in exchange for additional revenue. This market-based mechanism incentivizes active participation by generators while ensuring that sufficient ramping capability is available to maintain system reliability under real-time operational uncertainty.

While large-scale generators have traditionally provided these services, there is growing interest in leveraging distribution systems as additional sources of flexibility. Distribution systems, once passive, are now emerging as vital flexibility providers due to the proliferation of distributed energy resources (DERs). This growing interest has spurred research into aggregation-based models that represent the collective capability of these diverse assets. Specifically, recent studies by \cite{li2025aggregate, gupta2025grid} demonstrate that active distribution networks can yield substantial flexibility for the upstream grid. By coordinating and aggregating heterogeneous DERs, including energy storage systems (ESS), electric vehicles (EV), and flexible loads, these networks act as aggregated virtual power plants that support the broader system’s operational needs.

From a mathematical perspective, power aggregation can be viewed as a projection of high-dimensional, device-level operational constraints onto a low-dimensional feasible region of net substation power injections. This projection characterizes the aggregate flexibility that a distribution system can reliably offer to the transmission grid \cite{wen2022aggregate}. However, as noted by \cite{chen2019aggregate, ozturk2024alleviating}, explicitly computing this feasible region is computationally intractable at scale due to the large number of heterogeneous DERs, temporal coupling across time periods, and network security constraints.

To address this challenge, several recent studies focus on constructing conservative inner approximations of the aggregate feasible region \cite{chen2019aggregate,ozturk2024alleviating,wen2025quantifying,chen2021leveraging,al2024efficient}. Such approximations provide disaggregation guarantees, ensuring that any aggregate power trajectory within the approximated region can be reliably mapped back to a feasible dispatch of individual DERs. For clarity, we call an aggregate trajectory \emph{disaggregable} if any power setpoint within the computed flexibility region can be disaggregated while satisfying all underlying DER-level and network-feasibility constraints over the horizon.

Representative studies include the work by Chen et al.~\cite{chen2019aggregate}, that introduces an inner-approximation framework for aggregating multi-period flexibility in unbalanced distribution systems with disaggregation guarantees. This work is later extended in \cite{chen2021leveraging} using a two-stage adaptive robust optimization framework to provide stronger feasibility guarantees with reduced conservativeness. Building on this line of work on inner approximations, \cite{al2024efficient} proposes linear programming–based methods to compute inner approximations of the Minkowski sum of EV flexibility sets, while \cite{na2024aggregating} further integrates aggregation with cost-minimization frameworks for non-industrial air-conditioning systems and EVs. As summarized in Table~\ref{tab:related_work}, existing literature primarily focuses on state-coupled constraints but does not explicitly account for ramp-rate limits, making them difficult to apply to ramp-limited resources. 

Meanwhile, a growing number of generation and load-side resources are being integrated into power systems, driven by the expansion of data centers, artificial intelligence workloads, and other electricity-intensive applications \cite{gooding2024microsoft,cha2025potential}. Several of these resources, including smaller-scale or distributed deployments of fuel cells, combined heat and power (CHP) units \cite{alipour2019real}, and small modular reactors (SMRs) \cite{michaelson2021review}, exhibit inherent ramp-rate limitations. Therefore, there is need of aggregation schemes, where the ramp-limits are explicitly accounted.
 
To address this gap, we first develop a \textit{ramping-aware} flexibility aggregation framework that explicitly integrates ramp-rate constraints into the inner-box approximation. While this ensures disaggregation guarantees for ramp-limited resources, the integration of ramping constraints leads to an overly conservative flexibility envelope, as the feasible power range at each interval is constrained by the previous state. 
To overcome this conservativeness, we propose an \textit{enhanced} formulation that introduces new decision variables, referred to as ``pre-ramping'' decisions at interval $t-1$, which proactively reposition resource operating points to expand the feasible power range available at interval $t$. This approach helps partially decouple the current flexibility setpoint from the previous interval. The pre-ramping variables are optimized to enable delivering maximum ramping capability when required. As summarized in Table~\ref{tab:related_work}, this mechanism effectively enlarges the aggregate flexibility envelope while strictly preserving disaggregation guarantees and grid feasibility.

To the best of our knowledge, this is the first grid-aware aggregation framework that account for explicit ramping constraints with provable disaggregation guarantees. While existing work in Huang et al. \cite{huang2023research} address ramping through iterative boundary-shrinkage, they do not guarantee disaggregation. Moreover, the proposed pre-ramping scheme enhances the flexibility region compared to the baseline model. This performance gain is quantitatively validated in Section~\ref{sec:numerical_validation}.

The main contributions of this paper are listed below.
\begin{enumerate}
    \item We propose a flexibility aggregation framework that explicitly accounts for ramp-rate constraints of resources and guarantees the existence of a feasible unit-level disaggregation for any aggregate trajectory within the constructed envelope. It is referred to as the \textbf{Ramping-aware baseline model}.
    \item We introduce pre-ramping decisions that reposition resource operating points in advance to enlarge the aggregate flexibility envelope, while preserving disaggregation guarantees and network feasibility. It is referred to as the \textbf{Pre-ramped enhanced model}.
    \item We apply the proposed aggregation framework to a FRP market setting and quantitatively evaluate its effectiveness in enhancing practically deliverable flexibility.
\end{enumerate}

The paper is organized as follows: Section~\ref{sec:Old_Formulation} presents the baseline flexibility envelope formulation without pre-ramping. Section~\ref{sec:New_Formulation} introduces the proposed pre-ramped flexibility envelope model and details its mathematical formulation. Section~\ref{sec:numerical_validation} provides numerical case studies to validate the effectiveness of the proposed framework. Finally, Section~\ref{sec:conclusion} concludes the paper and discusses key findings.

\section{Ramping-Aware Flexibility Aggregation: Baseline Model}
\label{sec:Old_Formulation}
We consider a power distribution system with multiple distributed generators (DGs) and ESSs. The objective is to aggregate the combined flexibility from DGs and ESSs at the grid connection point (GCP), i.e., at the transmission--distribution interconnection. We next present an aggregation model that takes ramping constraints into account. It is referred to as baseline model and described below. 

The baseline model constructs an inner-box approximation of the aggregate flexibility at the GCP. The model explicitly accounts for resource-level operational constraints and distribution-network constraints, while ensuring temporal feasibility through ramping and time-coupled constraints.
Specifically, aggregate flexibility is represented by a pair of upper and lower feasible power trajectories at the GCP over the considered time horizon. These trajectories define a time-coupled flexibility envelope such that any aggregate power trajectory lying between them is disaggregable into a feasible dispatch of individual resources. The proposed formulation aims to determine the largest possible envelope by jointly optimizing the upper and lower trajectories subject to resource-level operating limits, inter-temporal ramping constraints, energy-coupling constraints, and distribution-network feasibility.

\subsection{Optimization Problem}
\label{sec:baseline_model}
We next present the objective function as well as resource and network constraints.
\subsubsection{Objective Function}
The objective maximizes the aggregate flexibility area at the GCP over the considered time horizon. This is achieved by maximizing the difference between the upper and lower feasible power trajectories. Using the symbols from the nomenclature, it is
\begin{align}
    \max \sum_{t} \left( p^{\text{GCP},\wedge}_{t} - p^{\text{GCP},\vee}_{t} \right)\cdot \Delta t.
    \label{eq:grid_objective}
\end{align}
\subsubsection{Resources Constraints}
The resource constraints define feasible upper and lower operating trajectories for generators and ESSs. Generator power outputs are constrained within their operating limits for both the upper and lower envelope trajectories. These constraints are enforced as follows:
\begin{subequations}
\label{eq:gen_power_limit_old}
\begin{align}
P^{\min}_g \le\;& p^{s}_{g,t} \le P^{\max}_g,
&& \forall g \in \mathcal{G},\;t \in \mathcal{T},\;s\in\{\wedge,\vee\}, \label{eq:gen_power_limit_old1} \\
Q^{\min}_g \le\;& q^{s}_{g,t} \le Q^{\max}_g,
&& \forall g \in \mathcal{G},\;t \in \mathcal{T},\;s\in\{\wedge,\vee\}, \\
p^{\wedge}_{g,t} \ge\;& p^{\vee}_{g,t},
&& \forall g \in \mathcal{G},\;t \in \mathcal{T}.
\end{align}
\end{subequations}

As discussed earlier, we consider generator ramping constraints between the subsequent time periods are defined as:%
\begin{subequations}
\label{eq:gen_ramp_old}
\begin{align}
p^{s}_{g,t} - p^{s}_{g,t-1} &\leq R^{\uparrow}_g, && \forall g \in \mathcal{G},\;t\ge2, s\in\{\wedge,\vee\}, \\
p^{s}_{g,t-1} - p^{s}_{g,t} &\leq R^{\downarrow}_g, && \forall g \in \mathcal{G},\;t\ge2, s\in\{\wedge,\vee\},\\
p^{\wedge}_{g,t} - p^{\vee}_{g,t-1} &\leq R^{\uparrow}_g, && \forall g \in \mathcal{G},\;t\ge2, \label{eq:gen_ramp1_m} \\
p^{\vee}_{g,t-1} - p^{\wedge}_{g,t} &\leq R^{\downarrow}_g, && \forall g \in \mathcal{G},\;t\ge2,  \\
p^{\vee}_{g,t} - p^{\wedge}_{g,t-1} &\leq R^{\uparrow}_g, && \forall g \in \mathcal{G},\;t\ge2,  \\
p^{\wedge}_{g,t-1} - p^{\vee}_{g,t} &\leq R^{\downarrow}_g, && \forall g \in \mathcal{G},\;t\ge2. \label{eq:gen_ramp4_m}
\end{align}
\end{subequations}

These ramp constraints are expressed for worst-case cross transitions, i.e., from lower to upper envelopes and vice-versa, as well as for upper and lower envelopes. 
These constraints ensure that transitions between the upper and lower envelopes remain achievable under finite ramp-rate limits. Specifically, the cross-corner constraints in \eqref{eq:gen_ramp_old} capture the most restrictive transitions, such as moving from the lower bound at $t-1$ to the upper bound at $t$. By bounding these worst-case power changes, the formulation guarantees that any interior trajectory can be disaggregated into a feasible resource-level schedule.

Similarly, we include constraints on ESS power outputs:
\begin{subequations}
\label{eq:ess_power_cons}
\begin{align}
-P^{\max}_e \leq &\; p^{s}_{e,t} \leq P^{\max}_e, && \forall e \in \mathcal{E},\; t \in \mathcal{T},\; s\in\{\wedge,\vee\}, \label{eq:ess_power_upper}\\
p^{\wedge}_{e,t} &\geq \; p^{\vee}_{e,t}, && \forall e \in \mathcal{E}, \;t \in \mathcal{T}.  \label{eq:ess_power_limit}
\end{align}

Then, the state-of-charge (SoC) dynamics and the constraints on the SoC are expressed as
\begin{align}
e^{s}_{e,t} &= \kappa_e \cdot e^{s}_{e,t-1} - p^{s}_{e,t-1} \cdot \Delta t, \;\forall e \in \mathcal{E},\; t\ge2,\; s\in\{\wedge,\vee\},   \label{eq:soc_conv1_old}  \\
&e^{\wedge}_{e,t} \ge E^{\text{init}}_e - \Delta t \sum_{\tau=1}^{t-1} p^{\vee}_{e,\tau},   \; \qquad \forall e \in \mathcal{E},t\ge2,  \label{eq:soc_conv1} 
\\
&e^{\vee}_{e,t} \le  E^{\text{init}}_e - \Delta t \sum_{\tau=1}^{t-1} p^{\wedge}_{e,\tau}, \; \qquad \forall e \in \mathcal{E},t\ge2,   \label{eq:soc_conv2}
\\
&E^{\min}_e \leq e^{s}_{e,t} \leq E^{\max}_e, \; \;\;  \forall e \in \mathcal{E},\; t \in \mathcal{T},\; s\in\{\wedge,\vee\}.
\label{eq:soc_min_old}
\end{align}
\end{subequations}

In \eqref{eq:soc_conv1_old}, $\kappa_e$ denotes the storage efficiency factor that models energy loss over time. We assume ideal charging and discharging efficiency\footnote{The losses can be taken into account by adding an equivalent resistance in series with the ideal ESS as implemented in \cite{gupta2023optimal}, which can be embedded in the LinDistFlow model.}, i.e., $\kappa_e = 1$, as in \cite{chen2019aggregate, chen2021leveraging}, to simplify the disaggregation analysis. Constraints \eqref{eq:soc_conv1} and \eqref{eq:soc_conv2} ensure that, even if the ESS follows the most energy-accumulating (for \( e^{\wedge}_{e,t} \)) or energy-depleting (for \( e^{\vee}_{e,t} \)) trajectory up to time \(t\), the SoC remains within energy limits determined by its initial energy state. Equation \eqref{eq:soc_min_old} ensures that the SoC stays within its admissible energy limits.

Finally, the initial conditions for the upper and lower trajectories are given by:
\begin{subequations}
\begin{align}
p^{\wedge}_{g,0} &= p^{\vee}_{g,0} = P^{\text{init}}_{g}, && \forall g \in \mathcal{G},\\ 
e^{\wedge}_{e,0} &= e^{\vee}_{e,0} = E^{\text{init}}_{e}, && \forall e \in \mathcal{E},\\
p^{\wedge}_{e,0} &= p^{\vee}_{e,0} = 0, && \forall e \in \mathcal{E}. 
\end{align}
\end{subequations}

\subsubsection{Network Constraints}
We model the network constraints using the linearized DistFlow model (LinDistFlow) \cite{baran1989network}. 

Let $B \in \mathbb{R}^{|\mathcal{N}|\times|\mathcal{L}|}$ denote the node--branch incidence matrix, 
$\boldsymbol{r},\boldsymbol{x} \in \mathbb{R}^{|\mathcal{L}|}$ collect line resistances and reactances, and define
$R := \mathrm{diag}(\boldsymbol{r}), X := \mathrm{diag}(\boldsymbol{x}).$
Let $A$ be the network incidence matrix and define matrices $H_P = A^{-1}RA^{-\top}$ and $H_Q = A^{-1}XA^{-\top}$. Then, the voltage magnitudes are given as%
\begin{subequations}
\begin{align}
     \boldsymbol{u}^s_t 
        = &\boldsymbol{1} + H_P\,\boldsymbol{p}^{\mathrm{inj},s}_t 
                      + H_Q\,\boldsymbol{q}^{\mathrm{inj},s}_t, &&t \in \mathcal{T},\;s\in\{\wedge,\vee\}
    \label{eq:lindistflow_vector}\\
    &\underline{\boldsymbol{u}} \;\leq\; \boldsymbol{u}^s_t \;\leq\; \overline{\boldsymbol{u}}, &&t \in \mathcal{T},\;s\in\{\wedge,\vee\}.
    \label{eq:voltage_limits}
\end{align}
\end{subequations}
Constraint \eqref{eq:voltage_limits} enforces lower and upper bounds on the squared nodal voltages, ensuring that voltage magnitudes remain within their admissible limits for all time periods and for both the upper and lower flexibility envelope trajectories.

The power injections and branch flows for all $t \in \mathcal{T}$ and
$s \in \{\wedge,\vee\}$ are given as
\begin{subequations}
\begin{align}
    \boldsymbol{p}^{\mathrm{inj},s}_t &= \boldsymbol{p}^s_{g,t} + \boldsymbol{p}^s_{e,t} + \boldsymbol{p}^{\text{PV}}_t - \boldsymbol{p}^{\text{load}}_t,
   \label{eq:P_injection_old}\\
    \boldsymbol{q}^{\mathrm{inj},s}_t
    &= \boldsymbol{q}^s_{g,t} - 
    \boldsymbol{q}^{\text{load}}_t,
    \label{eq:Q_injection_old}\\
        B\,\boldsymbol{P}^s_t &= \boldsymbol{p}^{\mathrm{inj},s}_t,
    \label{eq:branch_nodal_balance_P} \\
    B\,\boldsymbol{Q}^s_t &= \boldsymbol{q}^{\mathrm{inj},s}_t.
    \label{eq:branch_nodal_balance_Q} 
\end{align}
\end{subequations}
where \(\boldsymbol{p}^s_{g,t}\) and \(\boldsymbol{p}^s_{e,t}\) define the generator and ESS injection vectors for each trajectory \(s\) and time \(t\). The terms \(\boldsymbol{p}^s_{g,t}\) and \(\boldsymbol{p}^s_{e,t}\) are given as \(\sum_{g\in G(n)} p^s_{g,t}\) and \(\sum_{e\in E(n)} p^s_{e,t}\), respectively, where  \(G(n)\) and \(E(n)\) denote the sets of generators and ESS units located at node \(n\).
Here, branch power-flows are denoted by $\boldsymbol{P}^s_t$ and $\boldsymbol{Q}^s_t$, and
$B$ is matrix relating the injections to the branch flows.
Constraints \eqref{eq:P_injection_old}–\eqref{eq:Q_injection_old} define these nodal injections as the aggregation of generator and ESS outputs at each node minus the exogenous nodal load, thereby linking resource-level decisions to the network power-flow model. 

Finally, at the GCP, the following constraints are imposed:
\begin{subequations}
\label{eq:GCP_old}
\begin{align}
    p^{\mathrm{GCP},s}_{t}
        &= \boldsymbol{1}^\top \boldsymbol{p}^{\mathrm{inj},s}_t,
        &&\forall t \in \mathcal{T},\ \forall s \in \{\wedge,\vee\},
    \label{eq:GCP_power_def}
    \\
    p^{\mathrm{GCP},\wedge}_{t}
        &\ge p^{\mathrm{GCP},\vee}_{t},
        &&\forall t \in \mathcal{T}.
    \label{eq:GCP_envelope_order}
\end{align}
\end{subequations}
Constraint \eqref{eq:GCP_power_def} defines the net active power exchanged at the GCP as the aggregation of nodal active-power injections, where $\boldsymbol{1}$ is a vector of ones. Constraint \eqref{eq:GCP_envelope_order} enforces the ordering between the upper and lower aggregate trajectories, ensuring a valid flexibility envelope.

\vspace{0.5em}

\subsubsection{Final formulation}
\label{sec:old_Final_formulation}
The optimization problem in \eqref{eq:grid_objective}--\eqref{eq:GCP_old} is linear and can be written in the following standard form.
\begin{subequations}  
\begin{align}
    \text{max}~& \mathbf{c}^\top \mathbf{x}\\
    \text{subject to:}~ & \mathbf{A}\mathbf{x}\leq \mathbf{b}\\
    & \mathbf{C}\mathbf{x}\leq \mathbf{d},
\end{align}
\end{subequations}
where $\mathbf{c}$, $\mathbf{b}$, $\mathbf{d}$, and $\mathbf{A}$, $\mathbf{C}$ are appropriate vectors and matrices. The variable $\mathbf{x}$ contains all the variables, i.e., $[\mathbf{p}^s_{g,t}, \mathbf{q}^s_{g,t}, \mathbf{p}^s_{e,t}, \mathbf{e}^s_{e,t}, \boldsymbol{u}^s_{e,t}, \forall t, s \in \{\wedge,\vee\}]$. 

\subsection{Conservativeness of Ramping-aware Formulation}
\label{sec:issue_ramping}
As will be demonstrated via the numerical validations in Section~\ref{sec:numerical_validation}, the above formulation with ramping constraints results in a substantially smaller flexibility region than ones without. This is due to the worst-case transitions from upper to lower envelopes and vice-verse as included in constraints \eqref{eq:gen_ramp1_m}--\eqref{eq:gen_ramp4_m}. To tackle this issue, we next propose a new model.
\section{Ramping-aware Enhanced Flexibility Aggregation Model using Pre-ramping Strategy}
\label{sec:New_Formulation}
In order to address the conservativeness identified in the baseline formulation, we extend the model in Section~\ref{sec:Old_Formulation} by introducing pre-ramping variables that enable resources to pre-set their outputs, thereby expanding the available ramping flexibility in subsequent time intervals.

\begin{figure}[!b]
  \centering
  \includegraphics[width=\columnwidth]{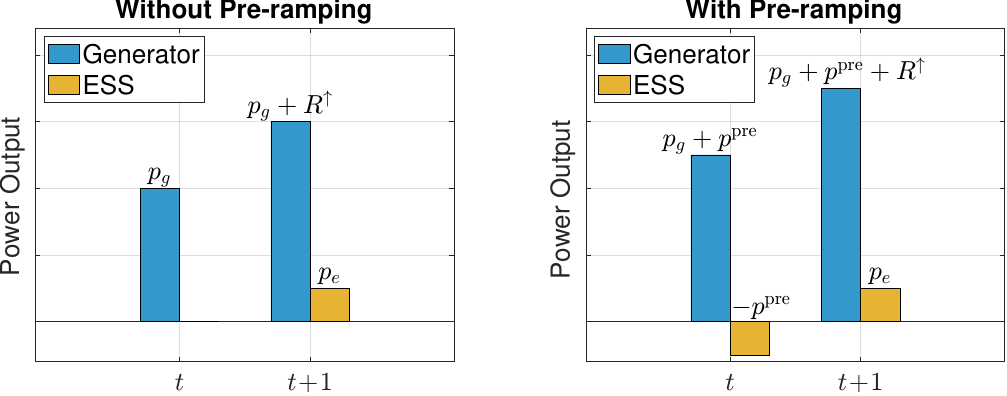}
  \vspace{-1.5em}
  \caption{Comparison of upward flexibility with and without pre-ramping. As shown, the upward flexibility increases from $p_g + R^{\uparrow} + p_e$  to $p_g + R^{\uparrow} + p_e + p^{\mathrm{pre}}$, where $p^{\mathrm{pre}}$ captures the additional ramp-up capability provided by pre-ramping.}
  \label{fig:pre_ramp_new}
\end{figure}

The conceptual example shown in Figure~\ref{fig:pre_ramp_new} illustrates the impact of pre-ramping on upward flexibility. In the \textit{first} scenario (left panel), no pre-ramping is applied. The generator can increase its output only up to its ramp-rate limit, while the ESS can contribute at most its available discharge capacity. As a result, the total output at time $t+1$ is limited to $p_g + R^\uparrow + p_e$.

In the \textit{second} scenario (right panel), pre-ramping is enabled. By pre-increasing the generator output and simultaneously pre-decreasing the ESS output at time $t$, the system maintains the same net output at the GCP. This re-positioning allows both units to fully exploit their upward ramping capability in the next interval, leading to a higher achievable aggregate output at time $t+1$, i.e., $p_g + R^{\uparrow} + p_e + p^{\mathrm{pre}}$, where the additional term $p^{\mathrm{pre}}$ captures the additional ramp-up capability provided by pre-ramping.

We next formally describe the enhanced model with pre-ramping variables. These variables are defined separately for generators and ESSs, and for the upper ($\wedge$) and lower ($\vee$) flexibility envelopes, as summarized in Table~\ref{tab:pre_ramp_interpretation}. All pre-ramping variables are defined as non-negative, with their physical interpretations enforced through the constraints. 

\begin{table}[!htbp]
\centering
\begin{minipage}{\columnwidth}
\centering
\caption{Interpretation of pre-ramping variables}
\label{tab:pre_ramp_interpretation}
\begin{tabular}{l l p{3.7cm} l}
\toprule
\textbf{Direction} & \textbf{Resource} & \textbf{Pre-Ramping Action} & \textbf{Variable} \\
\midrule
\multirow{2}{*}{Upward} 
  & Generator & Increase output & \( p^{\text{pre},\vee}_{g,t} \) \\
  & ESS       & Decrease output (charge more) & \( p^{\text{pre},\vee}_{e,t} \) \\
\midrule
\multirow{2}{*}{Downward} 
  & Generator & Decrease output & \( p^{\text{pre},\wedge}_{g,t} \) \\
  & ESS       & Increase output (discharge more) & \( p^{\text{pre},\wedge}_{e,t} \) \\
\bottomrule
\end{tabular}

\vspace{-0.5em}
\begin{flushleft}
\textit{\small Note: Each pre-ramping variable is associated with the envelope opposite to the direction of ramping. That is, ramp-up flexibility is prepared along the lower envelope (\( \vee \)), and ramp-down flexibility along the upper envelope (\( \wedge \)).}
\end{flushleft}
\end{minipage}
\vspace{-1.5em}
\end{table}

\subsection{Modified Constraints}

We define pre-ramped variables for generators and ESSs as $p^{\vee, \bullet}_{g,t} =  p^{\vee}_{g,t} +p^{\text{pre},\vee}_{g,t}$, $p^{\wedge, \bullet}_{g,t} = p^{\wedge}_{g,t} -p^{\text{pre},\wedge}_{g,t}$, and $p^{\vee, \bullet}_{e,t} =  p^{\vee}_{e,t} -p^{\text{pre},\vee}_{e,t}, p^{\wedge, \bullet}_{e,t} =  p^{\wedge}_{e,t} +p^{\text{pre},\wedge}_{e,t}$. The modified constraint are same as \eqref{eq:gen_ramp1_m}--\eqref{eq:gen_ramp4_m} but the generator and ESS variables are replaced by pre-ramped variables.
For generators, ramping feasibility is enforced by bounding all corner-to-corner transitions between the pre-ramped operating points at time $t\!-\!1$ and the envelope bounds at time $t$, which results in a set of eight linear ramping constraints; the complete set is reported in Appendix~\ref{app:preramp_disagg}. For ESS units, no ramp-rate limits are imposed; the baseline power and energy constraints retain the same structure and are evaluated using the pre-ramped variables.

\subsection{Additional Constraints to Ensure Feasibility and Coordination}

To ensure that the pre-ramped operating points remain physically realizable and compatible with both resource and network limits, additional feasibility and consistency constraints are introduced. These constraints guarantee that all upper and lower envelope trajectories, together with their associated pre-ramping actions, remain within the operational bounds of the generators, ESS units, and the distribution network. Moreover, these constraints are formulated to ensure that any trajectory contained within the resulting flexibility envelopes is disaggregable into individual resource schedules, as formally established in Appendix~\ref{app:preramp_disagg}. 

Along with preventing excessive pre-ramping actions that violate the energy boundaries of the ESS, the total pre-ramping energy over time must also respect the initial SoC. Specifically, the following constraints ensure that the cumulative upward and downward pre-ramping energy does not exceed the remaining chargeable or dischargeable energy, respectively:
\begin{subequations}
 \label{eq:pre_ramp_init_12}
\begin{align}
\sum_{t} p^{\text{pre},\wedge}_{e,t} \cdot \Delta t &\leq E^{\text{init}}_e - E^{\min}_e, && \forall e \in \mathcal{E},\\
\sum_{t} p^{\text{pre},\vee}_{e,t} \cdot \Delta t &\leq E^{\max}_e - E^{\text{init}}_e, && \forall e \in \mathcal{E}. \label{eq:pre_ramp_init_2}
\end{align}
\end{subequations}

It is worth noting that \( p^{\text{pre},\vee}_{e,t} \) serves to support upward flexibility by reducing the ESS output in advance. This effectively increases charging, which is only possible if sufficient room remains below the maximum energy capacity. Conversely, \( p^{\text{pre},\wedge}_{e,t} \) increases ESS output (discharging) to support downward flexibility, and is limited by the energy stored.

Pre-ramping is used to prepare future ramping capability and should not alter the net power injected at the GCP. In other words, from the perspective of the upstream grid, the GCP power must remain the same before and after applying the pre-ramp adjustments. This can be enforced by
\begin{subequations}
\label{eq:GCP_coordination_new}
\begin{align}
    p^{\text{GCP},\wedge}_{t} &= p^{\text{GCP},\wedge\bullet}_{t},
    && \forall t \in \mathcal{T}, \\
    p^{\text{GCP},\vee}_{t} &= p^{\text{GCP},\vee\bullet}_{t},
    && \forall t \in \mathcal{T}.    
\end{align}
\end{subequations}

Although the GCP invariance constraints \eqref{eq:GCP_coordination_new} already imply the coordination between generators and ESSs during pre-ramping, we explicitly impose the following balancing constraints to clearly reflect the physical coordination mechanism between resource types:
\begin{subequations}
\label{eq:pre_ramp_coordination_new}
\begin{align}
\sum_{g \in \mathcal{G}} p^{\text{pre},\wedge}_{g,t} &= \sum_{e \in \mathcal{E}} p^{\text{pre},\wedge}_{e,t},     && \forall t \in \mathcal{T}, \\ 
\sum_{g \in \mathcal{G}} p^{\text{pre},\vee}_{g,t} &= \sum_{e \in \mathcal{E}} p^{\text{pre},\vee}_{e,t},     && \forall t \in \mathcal{T}. 
\end{align}
\end{subequations}

\subsection{Final Formulation of the Pre-Ramped Model}

The final formulation of the proposed pre-ramped flexibility envelope model can be compactly expressed as follows. The key idea is that flexibility at the GCP is evaluated with respect to a baseline operating point, while pre-ramping actions are introduced to expand the aggregate flexibility envelope. Importantly, pre-ramping actions are defined on top of the baseline operating constraints rather than replacing them. Accordingly, both the baseline constraints and the associated pre-ramping constraints are jointly enforced to guarantee the feasibility of the expanded flexibility region.
\vspace{0.5em}

\subsubsection{Final formulation}
Similar to the baseline model, the optimization problem with pre-ramping variables in \eqref{eq:grid_objective}--\eqref{eq:pre_ramp_coordination_new} is linear and can be written in the following standard form.
\begin{subequations}
\begin{align}
    \text{max}~& \mathbf{c}^\top \mathbf{x}\\
    \text{subject to:}~ & \mathbf{A}\mathbf{x}\leq \mathbf{b}\\
    & \mathbf{C}\mathbf{x} = \mathbf{d}\\
    & \mathbf{A}^\bullet \mathbf{x}^\bullet \leq \mathbf{b}^\bullet\\
    & \mathbf{C}^\bullet\mathbf{x}^\bullet = \mathbf{d}^\bullet\\
    & \eqref{eq:pre_ramp_init_12}, \eqref{eq:GCP_coordination_new}, \eqref{eq:pre_ramp_coordination_new}.
\end{align}%
\label{eq:optimization_problem_final}%
\end{subequations}%
where $\mathbf{c}^\bullet$, $\mathbf{b}^\bullet$, $\mathbf{d}^\bullet$ and $\mathbf{A}^\bullet$, $\mathbf{C}^\bullet$ are appropriate vectors and matrices. The variable $\mathbf{x}^\bullet$ contains all the variables, i.e., $[\mathbf{p}^{s,\bullet}_{g,t}, \mathbf{q}^{s,\bullet}_{g,t}, \mathbf{p}^{s,\bullet}_{e,t}, p^{\text{pre},s}_{g,t}, p^{\text{pre},s}_{e,t}, \mathbf{e}^{s,\bullet}_{e,t}, \boldsymbol{u}^{s,\bullet}_{e,t}, \forall t, s \in \{\wedge,\vee\}].$

The baseline model can be interpreted as a special case of \eqref{eq:optimization_problem_final} by setting all pre-ramping variables to zero and removing the corresponding coordination constraints. It can be also observed that the optimization model in \eqref{eq:optimization_problem_final} is a linear program (LP) and can be solved efficiently by any LP solver.

\section{Numerical Validation}
\label{sec:numerical_validation}

\subsection{System Description and Simulation Setup}
\label{subsec:setup}

The proposed method is evaluated on the IEEE 33-bus radial distribution system (case33bw). We modify the system to include a ramp-constrained conventional generator, multiple ESSs, PV units, and time-varying loads. The detailed configurations and parameters of this system are provided in \cite{baran1989network}. The conventional generator is located at bus~5, while ESS units are installed at buses~10,~13,~14, and~24; PV units and loads are assumed to be distributed across multiple buses in the system.

The test system is modeled with a base power of 10~MVA and a base voltage of 12.66~kV. To maintain power quality, node voltage magnitudes are strictly limited to the range of $[0.95, 1.05]$~p.u. The dispatchable resources are characterized by a maximum capacity of 215~kW, a minimum of 80~kW, and a ramp-rate limit of 100~kW/h \cite{alipour2019real}. ESS units are assumed to have a power rating of 12.5~kW and an energy capacity of 50~kWh, with the initial state of charge set to 50\% of the rated capacity. Time-series profiles for PV generation and nodal loads are adopted from the smart grid test setup in \cite{gupta2025grid}.

Three case studies are considered. 
\begin{itemize}
    \item Case~I presents results on aggregate flexibility envelopes via area maximization.
    \item Case~II presents cost-optimal flexibility envelopes considering ramping products through net operational cost minimization under FRP markets.
    \item  Case~III presents a robust formulation accounting for uncertainty of PV and load forecasts.
\end{itemize}

For the economic evaluation in Case II, we consider hourly energy prices based on 2024 average CAISO locational marginal prices (LMPs), which ranges from \$5.8/\text{MWh} to \$66.7/\text{MWh}, with an average of \$32.4/\text{MWh} \cite{caiso_oasis}. Reserve capacity prices are set to \$20/\text{MW} for both upward and downward reserves, while compensation for providing FRPs is set to \$5.44/\text{MW} \cite{feng2022frequency, zhang2015impact}. The marginal generation cost of dispatchable units is assumed to be \$14.5/\text{MWh} \cite{alipour2019real}.

The simulation horizon spans 24 hours with an hourly resolution, but the proposed formulation is not tied to a specific dispatch interval. Under tighter real-time operational windows (e.g., 5 or 15 minutes), ramp-rate constraints become increasingly binding, which can substantially limit the deployable flexibility of ramp-constrained distributed energy resources.

\subsection{Case I: Aggregate Flexibility Envelopes via Area Maximization}
\label{subsec:area}
\begin{figure}[!t]
  \centering
  \begin{subfigure}{\columnwidth}
    \centering
    \includegraphics[width=0.99\linewidth]{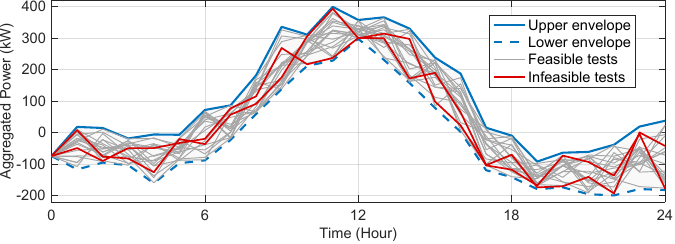}
    \vspace{-1.7em}
    \caption{Model without ramping constraints.}
    \label{fig:infeasible_wo_ramp}
  \end{subfigure}

  \begin{subfigure}{0.99\columnwidth}
    \centering
    \includegraphics[width=\linewidth]{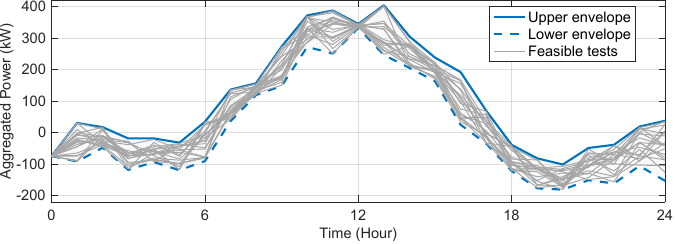}
    \vspace{-1.7em}
    \caption{Baseline model (ramping-aware).}
    \label{fig:infeasible_base}
  \end{subfigure}

  \begin{subfigure}{0.99\columnwidth}
    \centering
    \includegraphics[width=\linewidth]{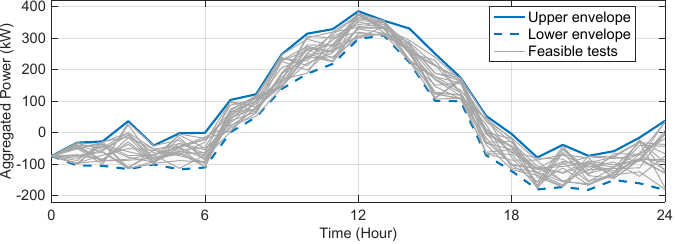}
    \vspace{-1.7em}
    \caption{Enhanced model with pre-ramping decisions.}
    \label{fig:infeasible_pre}
  \end{subfigure}

  \caption{Comparison of disaggregation feasibility for various GCP flexibility envelope models. Positive values represent power export from the distribution system to the upstream grid. (a) envelope neglecting ramp-rate constraints where red trajectories indicate infeasible disaggregation, (b) the baseline model considering ramping limits, and (c) the proposed enhanced model with pre-ramping decisions.} 
  \label{fig:envelopes_comparison_all}
\end{figure}

We examine the effectiveness of the proposed ramping-aware formulations in terms of feasibility with ramp constraints and in area within the aggregate flexibility envelopes. The optimized aggregated lower and upper envelopes are shown in Figure~\ref{fig:envelopes_comparison_all} for the model without ramping constraint (i.e., without eq.~\eqref{eq:gen_ramp_old}), the ramping-aware baseline model, and the enhanced model with pre-ramping decisions.

Figures~\ref{fig:envelopes_comparison_all}(a)-(c) show optimized upper and lower envelopes in solid and dashed blue. The trajectories in between the upper and lower envelopes are also included to show whether they can be disaggregated or not. The disaggregation feasibility of the constructed envelopes is verified through Monte Carlo simulations following the methodology established in \cite{chen2019aggregate}. Specifically, 1,000 vertex trajectories and 4,000 randomly sampled trajectories within the envelope are tested for each case. The trajectories in Figure~\ref{fig:envelopes_comparison_all} correspond to a representative subset of these Monte Carlo samples, selected solely for clarity of visualization.
As shown in Figure~\ref{fig:infeasible_wo_ramp}, when ramping constraints are neglected, several sampled trajectories inside the aggregate GCP envelopes cannot be disaggregated into feasible device-level schedules, as indicated by the infeasible trajectories (shown in red). This occurs because inter-temporal ramp-rate limits of dispatchable resources restrict the magnitude of feasible aggregate power changes between consecutive time intervals, even if the aggregate power remains within the envelope bounds.
By explicitly accounting for these ramping constraints, the ramping-aware baseline model in Figure~\ref{fig:infeasible_base} eliminates such infeasible trajectories, ensuring that all tested trajectories within the envelope are deliverable. Moreover, the enhanced model with pre-ramping decisions in Figure~\ref{fig:infeasible_pre} further expands the aggregate flexibility envelope by strategically re-positioning resources in advance, thereby enabling larger feasible power variations at the GCP. Importantly, all tested trajectories are found to be feasible, confirming that every aggregate power trajectory within the proposed envelopes can be disaggregated into device-level schedules. 

\begin{table}[!htbp]
\centering
\begin{threeparttable}
\caption{Impact of ESS Capacity and Placement on Aggregate Flexibility Envelope Area
(per-unit ESS ratings; four total ESS units).}
\label{tab:case1_envelope}
\begin{tabular}{c 
>{\centering\arraybackslash}p{1.7cm}
>{\centering\arraybackslash}p{1.7cm}
>{\centering\arraybackslash}p{2.2cm}}
\toprule
\textbf{ESS Rating} &
\multicolumn{3}{c}{\textbf{Envelope Area (kWh)}} \\
\textbf{(kW / kWh)} &
\textbf{w/o Ramping} &
\textbf{Baseline} &
\textbf{Pre-ramp} \\
\midrule
12.5 / 50   & 3324.90 & 2490.96 & 2619.90\,(+5.2\%) \\
12.5 / 50\tnote{$\dagger$} & 3329.71 & 2500.00 & 2635.00\,(+5.4\%) \\
25 / 100   & 3391.52 & 2557.57 & 2786.52\,(+9.0\%) \\
37.5 / 150 & 3457.62 & 2624.13 & 2952.62\,(+12.5\%) \\
62.5 / 250 & 3589.58 & 2756.09 & 3284.58\,(+19.2\%) \\
\bottomrule
\end{tabular}

\begin{tablenotes}
\small
\item[$\dagger$] ESS units are relocated to buses 5--8, near the generator.
\end{tablenotes}
\end{threeparttable}
\end{table}

Table~\ref{tab:case1_envelope} compares the aggregate flexibility envelope area at the GCP under different ESS capacity and placement scenarios. As shown in Table~\ref{tab:case1_envelope}, the flexibility gains enabled by pre-ramping grow significantly as ESS capacity increases, highlighting the key role of coordinated pre-ramping actions between the generator and ESS units. By proactively adjusting their operating points, the proposed formulation relaxes inter-temporal ramping constraints and unlocks additional aggregate flexibility. In contrast, the baseline formulation exhibits only modest improvements with increasing ESS capacity, as ramp-rate limits remain binding without pre-ramping.

The impact of resource placement is also examined. For the same ESS capacity, placing storage units on buses closer to the generator expands the flexibility envelope, as shown in the last row of Table~\ref{tab:case1_envelope}. This suggests that reduced electrical distances between dispatchable units can mitigate network-induced limitations and better facilitate the coordinated ramping required for envelope enlargement.

\subsection{Case II: Cost-Optimal Operation under FRP Markets}
\label{subsec:frp}
This case study investigates the economic implications of
the proposed pre-ramping formulation when the distribution
system participates in energy, reserve, and FRP markets.
In contrast to Case ~I, which focuses on maximizing the aggregate flexibility envelope, this case also optimizes a reference (base) trajectory at the GCP, representing the scheduled operating point of the distribution system.

Accordingly, three aggregate power trajectories are optimized simultaneously: an upper trajectory $p^{\mathrm{GCP},\wedge}_t$, a lower trajectory $p^{\mathrm{GCP},\vee}_t$, and a base trajectory $p^{\mathrm{GCP},-}_t$. The base trajectory is constrained to lie within the flexibility envelope,
\begin{equation}
p^{\mathrm{GCP},\vee}_t \;\le\; p^{\mathrm{GCP},-}_t \;\le\; p^{\mathrm{GCP},\wedge}_t, 
\qquad \forall t \in \mathcal{T},
\end{equation}
ensuring that both upward and downward flexibility are deliverable relative to the scheduled dispatch.

Based on these trajectories, the available upward and downward reserve capacities are defined as
\begin{equation}
R^{\mathrm{cap},\uparrow}_t = p^{\mathrm{GCP},\wedge}_t - p^{\mathrm{GCP},-}_t,
\;
R^{\mathrm{cap},\downarrow}_t = p^{\mathrm{GCP},-}_t - p^{\mathrm{GCP},\vee}_t.
\end{equation}
In addition, FRP quantities capture the deliverable ramping capability between consecutive time periods:
\begin{equation}
R^{\mathrm{frp},\uparrow}_t = p^{\mathrm{GCP},\wedge}_{t+1} - p^{\mathrm{GCP},-}_t,
\;
R^{\mathrm{frp},\downarrow}_t = p^{\mathrm{GCP},-}_t - p^{\mathrm{GCP},\vee}_{t+1}.
\end{equation}
These definitions, applied for all $t \in \mathcal{T}$, ensure that reserve and ramping services are consistent with inter-temporal ramp-rate constraints.

The objective function minimizes the net operating cost of the distribution system over the scheduling horizon:
\begin{equation}
\label{eq:new_obj}
\min \sum_t \Big( C_t^{\text{energy}} - R_t^{\text{reserve}} - R_t^{\text{FRP}} \Big).
\end{equation}
Here, $C_t^{\text{energy}}$ corresponds to the operating cost of the base trajectory and includes the energy procurement or export cost at the grid connection point, valued at energy prices, together with the generation cost of dispatchable generators. 

The revenue terms consist of compensation for reserve capacity and flexible ramping services. In particular, $R_t^{\text{reserve}}$ is determined by applying the reserve capacity price to the available upward and downward reserve margins, while $R_t^{\text{FRP}}$ is computed by valuing the deliverable ramping capacities using the FRP price defined in the market setup. This formulation directly links the economic incentives to the deliverable flexibility characterized by the aggregate power trajectories.

All resource-level operating constraints, ramping limits, and distribution-network constraints remain consistent with the flexibility envelope formulation in Section~\ref{sec:Old_Formulation} and Section~\ref{sec:New_Formulation}. By jointly optimizing the base dispatch and the flexibility envelope, the proposed model enables the distribution system to economically schedule its resources while simultaneously maximizing the value of reserve and ramping services.

\begin{figure}[!htbp]
  \centering
  \begin{subfigure}{\columnwidth}
    \centering
    \includegraphics[width=\linewidth]{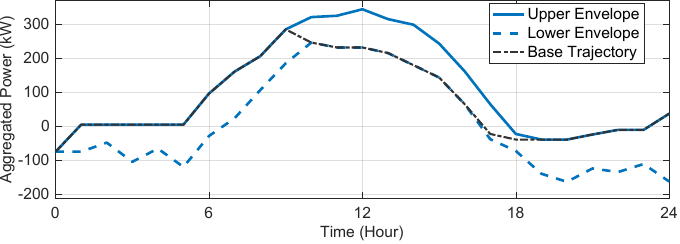}
    \vspace{-1.7em}
    \caption{Baseline model (ramping-aware)}    
    \label{fig:gcp_base}
  \end{subfigure}

  \begin{subfigure}{\columnwidth}
    \centering
    \includegraphics[width=\linewidth]{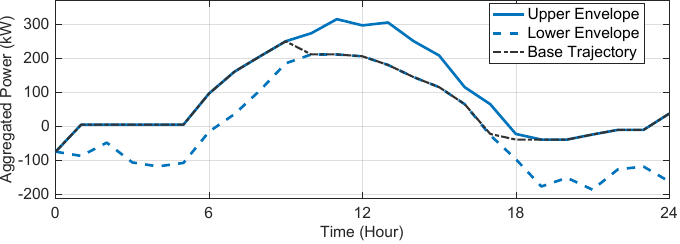}
    \vspace{-1.7em}
    \caption{Enhanced model with pre-ramping decisions}
    \vspace{-0.5em}
    \label{fig:gcp_pre}
  \end{subfigure}

  \caption{Aggregate power trajectories at the GCP in Case II. The black line denotes the base (reference) dispatch.}
  \vspace{-0.5em}
  \label{fig:envelopes_FRP}
\end{figure}

\begin{figure}[!htbp]
  \centering
  \begin{subfigure}{\columnwidth}
    \centering
    \includegraphics[width=\linewidth]{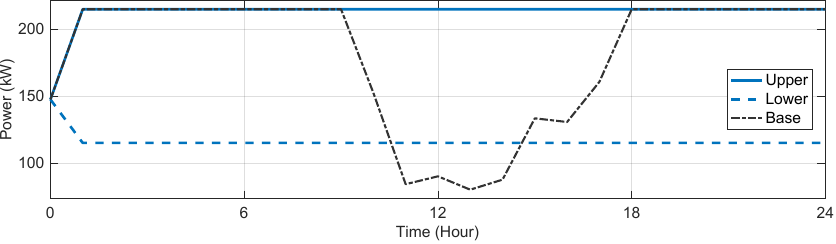}
    \vspace{-1.7em}
    \caption{Baseline model (ramping-aware)}    
    \label{fig:gen_base}
  \end{subfigure}

  \begin{subfigure}{\columnwidth}
    \centering
    \includegraphics[width=\linewidth]{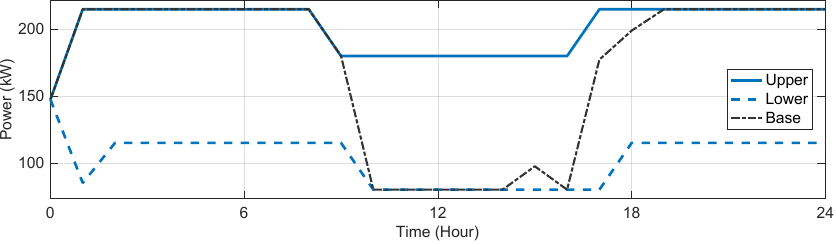}
    \vspace{-1.7em}
    \caption{Enhanced model with pre-ramping decisions}

    \label{fig:gen_pre}
  \end{subfigure}

  \caption{Generator output trajectories in Case II corresponding to the aggregate trajectories at the GCP. The black line denotes the base (reference) dispatch.}
  \label{fig:envelopes_FRP_Gen}
\end{figure}

\begin{table}[!htbp]
\centering
\caption{Economic Results: Comparison of Baseline and Pre-ramp Cases (per-unit ESS ratings; four ESS units in total)}
\label{tab:economic_results}
\fontsize{8.5}{10}
\setlength{\tabcolsep}{8.5pt}
\begin{tabular}{l c r r r}
\toprule
\textbf{ESS Rating} & \textbf{Case} & \textbf{$C^{\text{energy}}$} & \textbf{$R^{\text{res+frp}}$} & \textbf{Objective} \\
\textbf{(kW / kWh)} & & \textbf{(\$)} & \textbf{(\$)} & \textbf{(\$)} \\
\midrule
12.5 / 50   & Baseline  & 42.52  & 63.60  & -21.08 \\
            & Pre-ramp  & 41.23  & 64.36  & -23.14 \\
            & \textbf{Diff} & \textbf{-1.29} & \textbf{+0.76} & \textbf{-2.06} \\
\midrule
37.5 / 150  & Baseline  & 9.44   & 67.37  & -57.94 \\
            & Pre-ramp  & 8.19   & 68.26  & -60.07 \\
            & \textbf{Diff.} & \textbf{-1.25} & \textbf{+0.89} & \textbf{-2.14} \\
\midrule
62.5 / 250  & Baseline  & -20.87 & 70.02  & -90.89 \\
            & Pre-ramp  & -22.09 & 70.91  & -93.00 \\
            & \textbf{Diff.} & \textbf{-1.22} & \textbf{+0.89} & \textbf{-2.11} \\
            \midrule
125 / 500  & Baseline  & -71.76 & 73.81  & -145.57 \\
            & Pre-ramp  & -71.01 & 83.63  & -154.64 \\
            & \textbf{Diff.} & \textbf{+0.75} & \textbf{+9.82} & \textbf{-9.07} \\
\bottomrule
\end{tabular}
\end{table}

Figure~\ref{fig:envelopes_FRP} shows the results for the cost-optimal operation under FRP markets. It shows the aggregated lower and upper envelopes  along with the GCP base trajectory. Observe that the base (reference) trajectory is contained between the upper and lower envelopes. Figure~\ref{fig:envelopes_FRP_Gen} 
shows the generator output trajectories corresponding to the aggregate GCP trajectories. With pre-ramping enabled (Figure~\ref{fig:envelopes_FRP_Gen}(b)), the relaxed ramping constraints are clearly reflected in a larger separation between the upper and lower envelopes. Under the baseline formulation (Figure~\ref{fig:envelopes_FRP_Gen}(a)), the separation remains fixed at 100~kW across all time periods due to binding ramp-rate constraints. The generator's base trajectory exhibits similar patterns in both cases and always satisfies physical ramping limits. The base trajectory is not necessarily confined within the generator-level upper and lower bounds, as these bounds are auxiliary constructs for defining the aggregate GCP flexibility envelope rather than strict limits on the base dispatch.

The quantitative comparison of the costs are summarized in Table~\ref{tab:economic_results}. A negative energy ($C^\text{energy}$) value represents exports. Regardless of the installed ESS capacity, the introduction of the proposed pre-ramping consistently improves economic performance. Specifically, pre-ramping yields both higher revenues from reserve and FRP provision and lower operating costs. These gains arise from the expansion of the feasible operating region, which enables the distribution system to exploit flexibility more effectively without requiring additional physical resources, thereby enhancing economic efficiency through flexibility reallocation rather than capacity expansion.

Across all tested capacities, the pre-ramping strategy consistently improves the objective value compared to the baseline. For smaller ESS ratings, the improvement remains relatively modest, as the storage is mainly used for energy cost reduction. However, as the ESS capacity increases, the economic gain grows significantly, driven by a sharp rise in reserve and FRP revenues ($R^{\text{res+frp}}$). This shows that while pre-ramping is effective even at small scales, its true value is realized in larger systems where the additional capacity can be fully shifted from energy arbitrage to providing high-value flexibility services.

Overall, the results show that the proposed pre-ramping formulation enlarges the feasible flexibility region and improves economic performance in energy, reserve, and FRP markets.

\subsection{Case III: Robust Envelope under Forecast Errors}
\label{subsec:robust}
The above formulations assume that the PV and load forecasts can be obtained with 100\% certainty which is not realistic. Thus, we next consider a robust version of the proposed scheme accounting for uncertain PV and load injections. Due to the linear model, robust constraints can efficiently added in the previous formulation as described below.

To ensure the disaggregation of the flexibility envelope under net-load forecast uncertainty, we incorporate a robust optimization framework. Specifically, the nodal net-load realizations are modeled by box uncertainty bounds in both load demand $\boldsymbol{\xi}_t^{L}$ and PV generation $\boldsymbol{\xi}_t^{PV}$:
\begin{align}
\boldsymbol{p}^{\text{net}}_t 
= (\boldsymbol{p}^{\text{load}}_t + \boldsymbol{\xi}^L_t) - (\boldsymbol{p}^{\text{PV}}_t + \boldsymbol{\xi}^{PV}_t),
\label{eq:robust_netload_model}
\end{align}
where $\boldsymbol{p}^{\text{load}}_t$ and $\boldsymbol{p}^{\text{PV}}_t$ are the forecasted values. The uncertainty terms reside within a box uncertainty set $\mathcal{U}_t$:
\begin{align}
\mathcal{U}_t = \{ (\boldsymbol{\xi}_t^L, \boldsymbol{\xi}_t^{PV}) : |\boldsymbol{\xi}_t^L| \le \bar{\boldsymbol{\xi}}^L_t, \,\, |\boldsymbol{\xi}_t^{PV}| \le \bar{\boldsymbol{\xi}}^{PV}_t \}.
\label{eq:uncertainty_set}
\end{align}
The maximum anticipated deviations are defined as $\bar{\boldsymbol{\xi}}_t^L = \alpha |\boldsymbol{p}^{\text{load}}_t|$ and $\bar{\boldsymbol{\xi}}_t^{PV} = \beta |\boldsymbol{p}_t^{PV}|$, where $\alpha$ and $\beta$ represent the forecast error percentages for load and PV, respectively.

Since the voltage model \eqref{eq:lindistflow_vector} is affine, the uncertainty impacts nodal voltages through the term $-H_P (\boldsymbol{\xi}^L_t - \boldsymbol{\xi}^{PV}_t)$. \footnote{Reactive power uncertainty can be handled analogously using the sensitivity matrix $H_Q$.} To account for worst-case voltage deviations within $\mathcal{U}_t$, the robust counterpart of the nominal voltage constraints \eqref{eq:voltage_limits} is:
\begin{subequations}
\label{eq:robust_reformulation}
\begin{align}
    \max_{(\boldsymbol{\xi}_t^L, \boldsymbol{\xi}_t^{PV}) \in \mathcal{U}_t} \left\{ \boldsymbol{u}^s_t - H_P (\boldsymbol{\xi}_t^L - \boldsymbol{\xi}^{PV}_t) \right\} \le \overline{\boldsymbol{u}}, \\
    \min_{(\boldsymbol{\xi}_t^L, \boldsymbol{\xi}_t^{PV}) \in \mathcal{U}_t} \left\{ \boldsymbol{u}^s_t - H_P (\boldsymbol{\xi}_t^L - \boldsymbol{\xi}^{PV}_t) \right\} \ge \underline{\boldsymbol{u}}.
\end{align}
\end{subequations}

By applying the analytical robust reformulation for box uncertainty, the deterministic tightened margins are derived as $\boldsymbol{\Delta u}^{\text{unc}}_t := |H_P| (\bar{\boldsymbol{\xi}}^L_t + \bar{\boldsymbol{\xi}}^{PV}_t)$. Consequently, the robust model is obtained by replacing the nominal constraints \eqref{eq:voltage_limits} and with the following tightened versions:
\begin{subequations}
\label{eq:voltage_robust}
\begin{align}
& \underline{\boldsymbol{u}} + \boldsymbol{\Delta u}^{\text{unc}}_t
\le \boldsymbol{u}^s_t 
\le \overline{\boldsymbol{u}} - \boldsymbol{\Delta u}^{\text{unc}}_t,
\label{eq:robust_voltage_limits} \\
& \underline{\boldsymbol{u}} + \boldsymbol{\Delta u}^{\text{unc}}_t
\le \boldsymbol{u}^{s\bullet}_t 
\le \overline{\boldsymbol{u}} - \boldsymbol{\Delta u}^{\text{unc}}_t,
\label{eq:robust_preramp_voltage_limits}
\end{align}
\end{subequations}
for all $t \in \mathcal{T}$ and $s\in\{\wedge,\vee\}$. This formulation ensures that the flexibility envelope remains feasible even under simultaneous worst-case fluctuations of load and renewable generation.

To investigate the impact of net-load forecast uncertainty on aggregated flexibility, the proposed robust formulation is evaluated under different forecast error levels. For simplicity, identical error bounds are assumed for load and PV generation ($\alpha=\beta$), and varied from 0\% to 10\%. For each case, flexibility envelopes are computed for both the baseline and pre-ramped models using the robust voltage constraints.

Table~\ref{tab:robust_results} summarizes the resulting aggregate envelope areas at the GCP. As the forecast error increases, the envelope area gradually decreases, reflecting the proactive reservation of voltage margins. Although pre-ramping consistently outperforms the baseline, its relative improvement declines from 5.2\% to 3.7\% at $\alpha=\beta=10\%$, as the tightened voltage limits reduce the available dispatch range for pre-ramping actions.

\begin{table}[!htbp]
\centering
\caption{Impact of Forecast Errors on Robust Flexibility Envelopes}
\label{tab:robust_results}
\begin{tabular}{c  >{\centering\arraybackslash}p{1.5cm} 
>{\centering\arraybackslash}p{1.5cm} 
>{\centering\arraybackslash}p{1.2cm}  
>{\centering\arraybackslash}p{1.2cm}}
\toprule
\textbf{Forecast} &
\multicolumn{2}{c}{\textbf{Envelope Area (kWh)}} &
\multicolumn{2}{c}{\textbf{Voltage Extremes (pu)}} \\
\textbf{Error (\%)} &
\textbf{Baseline} & \textbf{Pre-ramp} &
$\max |V|$ & $\min |V|$ \\
\midrule
0   & 2490.96 & 2619.90 & 1.0500 & 0.9500 \\
3   & 2488.77 & 2610.41 & 1.0468 & 0.9511 \\
5   & 2487.31 & 2601.31 & 1.0447 & 0.9518 \\
10  & 2473.72 & 2564.41 & 1.0403 & 0.9533 \\
\bottomrule
\end{tabular}
\end{table}

From a feasibility perspective, all constructed envelopes satisfy the voltage limits. Increasing uncertainty pushes the nodal voltage extrema $(\max |V|, \min |V|)$ further away from the operational bounds (1.05 and 0.95~p.u.), reflecting increasingly conservative robust margins. The robustness of the envelopes is further validated via Monte Carlo simulations, which confirm that randomly sampled aggregate trajectories remain disaggregable and voltage-feasible.

Overall, these results highlight a trade-off between robustness and flexibility: while uncertainty necessitates conservative voltage margins that reduce the envelope size, the proposed pre-ramping strategy consistently preserves additional deliverable flexibility under worst-case conditions.

\section{Conclusions}
\label{sec:conclusion}
This paper presented a ramping-aware flexibility aggregation framework for power distribution networks leveraging distributed energy resources (DERs). Since ramp constraints inherently reduce the available flexibility region, we introduced a pre-ramping strategy to mitigate this limitation. The proposed approach proactively coordinates ramp-constrained resources (e.g., generators) with ramp-unconstrained resources (e.g., energy storage systems), thereby expanding the flexibility region despite ramping limitations.

The framework was validated on the IEEE 33-bus benchmark network with multiple distributed generators and storage units. Simulations show that the pre-ramping model significantly enhances the flexibility region through proactive actions of energy storage units. Additionally, a cost-optimal formulation was developed to enable participation in multiple markets, including energy, reserve capacity, and flexible ramping products. Comparative analysis shows that the pre-ramping model yields higher economic benefits than the baseline approach. Finally, the formulation was extended to incorporate uncertainties in load and PV forecasts, confirming its robustness and effectiveness under uncertain operating conditions.

\appendices
\renewcommand{\theequation}{\thesection.\arabic{equation}}
\section{Proof of Disaggregation Guarantee for the Baseline Model}
\label{app:proof_disagg}
\setcounter{equation}{0}

Consider the baseline envelope solutions $\{p_{g,t}^{\wedge},p_{g,t}^{\vee}\}$ and $\{p_{e,t}^{\wedge},p_{e,t}^{\vee}\}$ that satisfy the baseline constraints in Section~\ref{sec:baseline_model}, including \eqref{eq:gen_power_limit_old}--\eqref{eq:ess_power_cons}. Let the corresponding GCP envelopes be denoted by $\{p_{t}^{\mathrm{GCP},\wedge},p_{t}^{\mathrm{GCP},\vee}\}$.

For any aggregate trajectory $\{p_{t}^{\mathrm{GCP},o}\}$ such that $p_{t}^{\mathrm{GCP},\vee}\le p_{t}^{\mathrm{GCP},o}\le p_{t}^{\mathrm{GCP},\wedge}$ for all $t \in \mathcal{T}$, define the auxiliary coefficient $\lambda_t\in[0,1]$ as
\begin{equation}
\begin{aligned}
\lambda_t &:=
\begin{cases}
\dfrac{p_{t}^{\mathrm{GCP},\wedge}-p_{t}^{\mathrm{GCP},o}}
      {p_{t}^{\mathrm{GCP},\wedge}-p_{t}^{\mathrm{GCP},\vee}},
& p_{t}^{\mathrm{GCP},\wedge}>p_{t}^{\mathrm{GCP},\vee},\\
0, & \text{otherwise},
\end{cases} \\
p_{t}^{\mathrm{GCP},o}
&= \lambda_t p_{t}^{\mathrm{GCP},\vee}
 + (1-\lambda_t)p_{t}^{\mathrm{GCP},\wedge}.
\end{aligned}
\label{eq:lambda_app}
\end{equation}
We construct a disaggregation by convex interpolation:
\begin{equation}
p_{g,t}^{o}:=\lambda_t p_{g,t}^{\vee}+(1-\lambda_t)p_{g,t}^{\wedge},\qquad
p_{e,t}^{o}:=\lambda_t p_{e,t}^{\vee}+(1-\lambda_t)p_{e,t}^{\wedge}.
\label{eq:unit_interp_app}
\end{equation}
By linearity of aggregation, \eqref{eq:lambda_app}--\eqref{eq:unit_interp_app}
and the definition of GCP power in \eqref{eq:GCP_power_def} imply
\begin{equation}
p_{t}^{\mathrm{GCP},o}
= \boldsymbol{1}^\top
\big(
\boldsymbol{p}^{o}_{g,t}
+ \boldsymbol{p}^{o}_{e,t}
+ \boldsymbol{p}^{\mathrm{PV}}_t
- \boldsymbol{p}^{\mathrm{load}}_t
\big),
\quad \forall t \in \mathcal{T}.
\end{equation}
Moreover, since generator/ESS power limits are interval constraints, $p_{g,t}^{o}$ and $p_{e,t}^{o}$ remain feasible for all $t$ as convex combinations of feasible endpoints.

Next, we verify ramp feasibility. For each $g$ and $t\ge2$, the baseline model imposes the cross-corner bounds \eqref{eq:gen_ramp_old}, which ensure that all four ``corners'' $(p_{g,t-1}^{a},p_{g,t}^{b})$ with $a,b\in\{\vee,\wedge\}$ satisfy the ramp limits. Since the ramp constraints define a convex set in $(p_{g,t-1},p_{g,t})$, any point within the rectangle $[p_{g,t-1}^{\vee},p_{g,t-1}^{\wedge}]\times[p_{g,t}^{\vee},p_{g,t}^{\wedge}]$ also satisfies the ramp limits. Since $(p_{g,t-1}^{o},p_{g,t}^{o})$ lies in this rectangle by construction, the realized trajectory $\{p_{g,t}^{o}\}$ satisfies the ramp-rate limits.

For ESS energy feasibility, define the realized SoC dynamics $e_{e,1}^{o}=E_{e}^{\mathrm{init}}$ and $e_{e,t}^{o}=e_{e,t-1}^{o}-\Delta t\,p_{e,t-1}^{o}$ for $t\ge2$:
\begin{equation}
e_{e,t}^{o}=E_{e}^{\mathrm{init}}-\Delta t\sum_{\tau=1}^{t-1}p_{e,\tau}^{o}.
\label{eq:soc_real_app}
\end{equation}
Since $p_{e,\tau}^{\vee}\le p_{e,\tau}^{o}\le p_{e,\tau}^{\wedge}$ for all $\tau$,
\eqref{eq:soc_real_app} yields the sandwich bound
\begin{equation}
E_{e}^{\mathrm{init}}-\Delta t\sum_{\tau=1}^{t-1}p_{e,\tau}^{\wedge}
~\le~
e_{e,t}^{o}
~\le~
E_{e}^{\mathrm{init}}-\Delta t\sum_{\tau=1}^{t-1}p_{e,\tau}^{\vee}.
\label{eq:soc_sandwich_app}
\end{equation}
The baseline energy-coupling constraints \eqref{eq:soc_conv1}--\eqref{eq:soc_conv2}
(and bounds \eqref{eq:soc_min_old}) ensure that the left-hand side of
\eqref{eq:soc_sandwich_app} is no smaller than $E_{e}^{\min}$ and the right-hand side
is no larger than $E_{e}^{\max}$, thus $E_{e}^{\min}\le e_{e,t}^{o}\le E_{e}^{\max}$.

Therefore, for any aggregate trajectory inside the baseline envelope, the construction \eqref{eq:unit_interp_app}--\eqref{eq:soc_real_app} produces a feasible disaggregation that respects generator power/ramp constraints and ESS power/SoC constraints, while exactly matching the aggregate trajectory at the GCP.

Finally, the disaggregation guarantee directly extends to LinDistFlow network constraints. Since nodal injections, branch flows, and squared voltages are affine functions of device-level injections under LinDistFlow, and voltage limits define a convex (feasible) set, any convex interpolation of two feasible envelope solutions at time $t$ remains feasible. Thus, all interior aggregate trajectories preserve network feasibility.

\section{Proof of Disaggregation Guarantee for the Pre-Ramped Model}
\label{app:preramp_disagg}
\setcounter{equation}{0}

Consider the pre-ramped envelope solutions $\{p_{g,t}^{\wedge},p_{g,t}^{\vee}\}$, $\{p_{e,t}^{\wedge},p_{e,t}^{\vee}\}$ and the associated pre-ramping variables $\{p_{g,t}^{\text{pre},\wedge},p_{g,t}^{\text{pre},\vee}\}$ and $\{p_{e,t}^{\text{pre},\wedge},p_{e,t}^{\text{pre},\vee}\}$ obtained from Section~\ref{sec:New_Formulation}. These variables satisfy all pre-ramped feasibility constraints, including \eqref{eq:pre_ramp_init_12}--\eqref{eq:pre_ramp_coordination_new}. Let the corresponding GCP envelopes be denoted by $\{p_{t}^{\mathrm{GCP},\wedge}, p_{t}^{\mathrm{GCP},\vee}\}$. By \eqref{eq:GCP_coordination_new}, pre-ramping does not change the net power at the GCP, i.e., $p_{t}^{\mathrm{GCP},s}=p_{t}^{\mathrm{GCP},s\bullet}$ for all $t\in \mathcal{T}$ and $s\in\{\wedge,\vee\}$.

For any aggregate trajectory $\{p_{t}^{\mathrm{GCP},o}\}$ such that $p_{t}^{\mathrm{GCP},\vee}\le p_{t}^{\mathrm{GCP},o}\le p_{t}^{\mathrm{GCP},\wedge}$ for all $t \in \mathcal{T}$, define $\lambda_t\in[0,1]$ as in \eqref{eq:lambda_app}. We construct a disaggregation by convex interpolation:%
\begin{equation}%
p_{g,t}^{o}:=\lambda_t p_{g,t}^{\vee}+(1-\lambda_t)p_{g,t}^{\wedge},\qquad
p_{e,t}^{o}:=\lambda_t p_{e,t}^{\vee}+(1-\lambda_t)p_{e,t}^{\wedge},
\label{eq:unit_interp_app_preramp}
\end{equation}
and similarly for the pre-ramped operating points,
\begin{equation}
p_{g,t}^{o\bullet}:=\lambda_t p_{g,t}^{\vee\bullet}+(1-\lambda_t)p_{g,t}^{\wedge\bullet},\qquad
p_{e,t}^{o\bullet}:=\lambda_t p_{e,t}^{\vee\bullet}+(1-\lambda_t)p_{e,t}^{\wedge\bullet}.
\label{eq:unit_interp_bullet_app_preramp}
\end{equation}

By linearity of aggregation, \eqref{eq:lambda_app} and \eqref{eq:unit_interp_app_preramp} with \eqref{eq:GCP_power_def} imply%
\begin{equation}
p_{t}^{\mathrm{GCP},o}
= \boldsymbol{1}^\top
\big(
\boldsymbol{p}^{o}_{g,t}
+ \boldsymbol{p}^{o}_{e,t}
+ \boldsymbol{p}^{\mathrm{PV}}_t
- \boldsymbol{p}^{\mathrm{load}}_t
\big),
\quad \forall t \in \mathcal{T}.
\end{equation}

Moreover, since generator/ESS power limits are interval constraints, $p_{g,t}^{o}$ and $p_{e,t}^{o}$ remain feasible for all $t$ as convex combinations of feasible endpoints. The same argument applies to $p_{g,t}^{o\bullet}$ and $p_{e,t}^{o\bullet}$ because the pre-ramped operating points are also enforced to be feasible in the pre-ramped model.

Next, we verify generator ramp feasibility. In the pre-ramped model, the physical transition from $t\!-\!1$ to $t$ is evaluated between the pre-ramped operating point at $t\!-\!1$ and the envelope dispatch at $t$, i.e., between $(p^{\bullet}_{g,t-1},p_{g,t})$. Ramping feasibility is enforced by bounding all corner-to-corner transitions between $\{p^{\vee,\bullet}_{g,t-1},p^{\wedge,\bullet}_{g,t-1}\}$ and $\{p^{\vee}_{g,t},p^{\wedge}_{g,t}\}$, which yields the following eight linear constraints:
\begin{subequations}
\label{eq:gen_ramp_new}
\begin{align}
p^{\wedge}_{g,t} - \bigl( p^{\vee}_{g,t-1} +p^{\text{pre},\vee}_{g,t-1}\bigr)
    &\leq R^{\uparrow}_g, && \forall g \in \mathcal{G},\;t\ge2,  \label{eq:gen_ramp1} \\
\bigl( p^{\vee}_{g,t-1} +p^{\text{pre},\vee}_{g,t-1}\bigr)- p^{\wedge}_{g,t}
    &\leq R^{\downarrow}_g, && \forall g \in \mathcal{G},\;t\ge2,   \\
p^{\vee}_{g,t} - \bigl(p^{\wedge}_{g,t-1} - p^{\text{pre},\wedge}_{g,t-1}\bigr)
    &\leq R^{\uparrow}_g, && \forall g \in \mathcal{G},\;t\ge2,   \\
\bigl(p^{\wedge}_{g,t-1} - p^{\text{pre},\wedge}_{g,t-1}\bigr) - p^{\vee}_{g,t}
    &\leq R^{\downarrow}_g, && \forall g \in \mathcal{G},\;t\ge2,  \label{eq:gen_ramp4}\\
p^{\wedge}_{g,t} - \bigl(p^{\wedge}_{g,t-1} - p^{\text{pre},\wedge}_{g,t-1}\bigr)
    &\leq R^{\uparrow}_g, && \forall g \in \mathcal{G},\;t\ge2, \label{eq:gen_ramp5} \\
p^{\vee}_{g,t} - \bigl(p^{\vee}_{g,t-1} + p^{\text{pre},\vee}_{g,t-1}\bigr)
    &\leq R^{\uparrow}_g, && \forall g \in \mathcal{G},\;t\ge2, \label{eq:gen_ramp6} \\
\bigl(p^{\wedge}_{g,t-1} - p^{\text{pre},\wedge}_{g,t-1}\bigr) - p^{\wedge}_{g,t}
    &\leq R^{\downarrow}_g, && \forall g \in \mathcal{G},\;t\ge2, \label{eq:gen_ramp7} \\
\bigl(p^{\vee}_{g,t-1} + p^{\text{pre},\vee}_{g,t-1}\bigr) - p^{\vee}_{g,t}
    &\leq R^{\downarrow}_g, && \forall g \in \mathcal{G},\;t\ge2. \label{eq:gen_ramp8}
\end{align}
\end{subequations}
Since these inequalities bound the ramp for all four corner pairs, the feasible set in $(p^{\bullet}_{g,t-1},p_{g,t})$ is convex. Hence, the pair $(p^{o\bullet}_{g,t-1},p^{o}_{g,t})$ also satisfies the generator ramp-rate limits.

For ESS units, the energy-feasibility argument follows the same structure as in Appendix~\ref{app:proof_disagg}. The only difference is that, in the pre-ramped model, the SoC dynamics are driven by the pre-ramped net power $p_{e,t}^{\bullet}$. Since the pre-ramped endpoint SoC trajectories corresponding to $\{\vee\bullet,\wedge\bullet\}$ are enforced to be feasible by the ESS energy and budget constraints, the same sandwich argument applies. Network feasibility under LinDistFlow is preserved by convex interpolation and therefore follows directly from Appendix~\ref{app:proof_disagg}.


\bibliographystyle{IEEEtran}
\bibliography{biblio.bib}

\end{document}